**Propagación de haces ultrasónicos en estructuras periódicas: verificación de la existencia de cristales de sonido en estructuras triangulares**

**Ultrasonic beam propagation in periodic structures: verifying the existence of sound crystals triangular structures**

**Enrique Orduña-Malea**
Universidad Politécnica de Valencia (UPV)
enorma@upv.es

**Resumen:** Cuando las ondas se propagan a través de un medio con elementos dispersores y estos además están colocados de forma periódica, como ocurre en las estructuras cristalinas, la dispersión múltiple nos lleva a un fenómeno conocido como estructuras en bandas. Esto significa que las ondas se pueden propagar en un cierto rango de frecuencias, siguiendo unas reglas de dispersión, mientras que en otros rangos de frecuencias la propagación es eliminada. Las primeras son llamadas bandas permitidas y las últimas bandas prohibidas. Este trabajo parte con el objetivo de verificar la existencia de bandas prohibidas de frecuencia en la zona ultrasónica ante estructuras triangulares formadas por cilindros (huecos y macizos) de diversos diámetros (8 y 16mm.). Los resultados indican que la estructura triangular presenta zonas de atenuación selectivas (huecos frecuenciales), obteniendo resultados similares a los predichos teóricamente para las dos direcciones principales de simetría (0º y 30º).

**Palabras clave:** Cristales de sonido; Haces ultrasónicos; Estructuras periódicas; Ultrasonidos; Medio líquido; Cilindros.

## 1. INTRODUCCIÓN

Dado un medio a través del cual pueden propagarse ondas elásticas y al que llamamos "huésped", si introducimos otro material que, en el caso de cristales de sonido, tiene una densidad mayor y por tanto con una mayor velocidad de las ondas elásticas a su través, se forma lo que llamamos "material dispersor". Estas intrusiones son de forma periódica de manera que lo que se produce es una distribución periódica de densidades (las del material dispersor).

Además, si esta disposición periódica de densidades se realiza de forma que simulen determinadas estructuras cristalinas podremos conseguir patrones de dispersión muy especiales y es más, incluso bandas de frecuencia para las cuales el sonido no va a ser transmitido independientemente del ángulo de incidencia con el que llegue a la estructura. La posibilidad de estudio de estos patrones de dispersión mediante modelos a escala es el principal objetivo del presente trabajo.

### 1.1. Objeto y antecedentes
Cuando un conjunto de ondas se propaga a través de un medio que contiene muchos elementos dispersores, cada onda va a ser dispersada por cada uno de estos elementos y las ondas dispersadas volverán a ser dispersadas por los otros elementos. Este proceso se repite estableciéndose un patrón recursivo infinito de redispersión entre dispersores.

La dispersión múltiple de ondas es responsable de un buen número de fenómenos, como pueden ser la modulación del sonido ambiente en el océano o el centelleo acústico de bancos de peces entre muchos otros.

Cuando las ondas se propagan a través de un medio con elementos dispersores y estos además están colocados de forma periódica como ocurre en las estructuras cristalinas, la dispersión múltiple nos lleva a un fenómeno conocido como estructuras en bandas. Esto significa que las ondas se pueden propagar en un cierto rango de frecuencias, siguiendo unas reglas de dispersión, mientras que en otros rangos de frecuencias la propagación es eliminada. Las primeras son llamadas bandas permitidas y las últimas bandas prohibidas.

En ciertas condiciones, la inhibición de la propagación de la onda ocurre para todas las direcciones de incidencia, en ese caso podemos decir que se ha producido un "hueco" completo en la banda de frecuencia, el cual se ha demostrado que ocurre cercano a la frecuencia del pico de la reflexión de Bragg.

La aparición de estos huecos en las bandas frecuenciales, donde la propagación está prohibida, fue estudiada primeramente para ondas electrónicas en sólidos, proporcionando la base para la comprensión de las propiedades de los conductores, semiconductores y aislantes. Posteriormente estos estudios fueron aplicados en el campo de la óptica a través de la teoría de la difracción y desde hace escasos años a la acústica, campo objeto del presente trabajo.

## 1.2. Objetivos.

El objetivo principal de este trabajo es el de analizar la dispersión del sonido, en la región ultrasónica, provocada por diversas estructuras periódicas sumergidas en agua, con el fin de verificar en la práctica la aparición de bandas de gran atenuación acústica, predichas según la llamada "teoría de la difracción" para, finalmente, relacionar los resultados obtenidos con elementos arquitectónicos.

Los objetivos específicos para llevar a cabo este propósito son los siguientes:
- Verificar la aparición de bandas de frecuencia atenuadas en la propagación de haces ultrasónicos a través de estructuras periódicas sumergidas en agua. El análisis se centra en estructuras triangulares dejando otras estructuras de interés (como las cuadradas o hexagonales) para futuros estudios.
- Analizar los efectos de atenuación en función del diámetro de las estructuras (en este caso cilindros), de su composición (hueca o maciza) y del grado de incidencia (0º y 30º).

## 2. ESTADO DE LA CUESTIÓN

### 2.1. Cristales de sonido.

En el caso de bandas permitidas y prohibidas a la propagación del campo acústico, las ondas deben encontrarse con una distribución periódica de la densidad y que ésta sea comparable a la longitud de onda del frente incidente.

Pese a que la teoría acerca de la atenuación acústica provocada por estructuras periódicas está hace tiempo estudiada, los trabajos experimentales son recientes.

Este tipo de estructuras reciben el nombre de cristales de sonido. Un cristal de sonido se puede denominar como aquel sistema en el que existe una distribución periódica de la densidad, ordenado según las dimensiones de la longitud de onda acústica con la que van a interactuar, de forma que tengan lugar los fenómenos de interferencia.

Una de las primeras observaciones de ésta índole fue la realizada para medir la atenuación acústica de una escultura exhibida en la fundación Juan March en Madrid, y que actualmente se encuentra en el campus de la UPV (figura 1), la cual parecía poseer una estructura rígida apropiada para los experimentos (Martínez Sala et al., 1995).

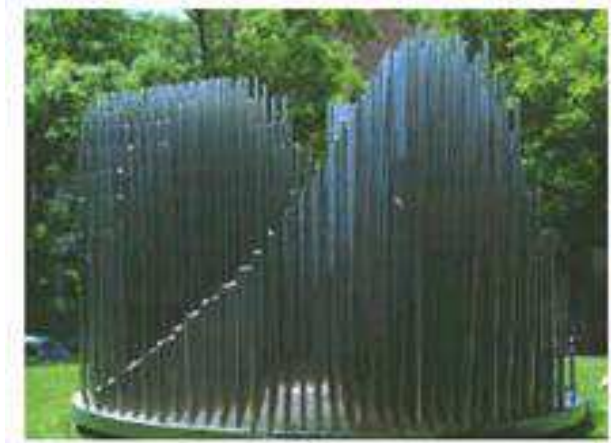

**Figura 1. Escultura de Eusebio Sempere**

Los autores obtuvieron un espectro de la atenuación del sonido que, posteriormente, fue comprobado mediante métodos computacionales. El objetivo de los autores era comprobar si el pico de atenuación que obtuvieron, cercano al primer pico de Bragg, podía ser considerado como el inicio de la formación del primer hueco en esta estructura.

Posteriormente se llevaron a cabo nuevas mediciones en la transmisión acústica a través de *arrays* de cilindros rígidos en dos dimensiones en medio aéreo (Rubio et al., 1997). Los autores demuestran las propiedades de la atenuación del sonido a lo largo de dos direcciones de alta simetría de la zona de Brillouin de los *arrays*. Además observan un peculiar efecto, denominado "bandas sordas", donde hay lugares del espectro en los cuales -teóricamente- debería propagarse el sonido, pero donde la propagación es prohibida debido a la simetría particular de las estructuras.

El estudio de la propagación de ondas acústicas/elásticas en sistemas periódicos puede ayudar a construir nuevos dispositivos como puedan ser filtros acústicos, guías de onda acústicas, transductores de nueva generación, pantallas acústicas, etc.

Los parámetros que controlan la aparición de bandas prohibidas son:

a) El tipo de simetría de la estructura.
b) El contraste de velocidad entre la onda propagándose por el material "huésped" y por el elemento dispersor.
c) El factor de llenado, definido como el ratio entre el volumen ocupado por cada dispersor respecto al total del volumen de la composición.
d) La topología:

   El fenómeno de múltiple dispersión se consigue mediante el uso de dos materiales, cada uno con diferentes velocidades de propagación.

   El componente de baja velocidad es tratado como el elemento dispersor mientras que el de mayor velocidad es denominado huésped.

   Se pueden distinguir dos tipos de topologías, según la forma del elemento dispersor:

- **Topología Cermet:** El material dispersor consiste en inclusiones aisladas, cada una de las cuales está completamente rodeada por el material huésped.
- **Topología Network:** El material dispersor está conectado y forma una continua "red de trabajo" a lo largo de toda la estructura.

Estudios teóricos (Economou y Sigalas, 1993) han demostrado que para ondas acústicas, la topología Cermet es, en general, más favorable para el desarrollo de huecos en contraste con las ondas electromagnéticas, para las cuales la topología Network es más favorable.

## 2.2. Bandas prohibidas de energía

La reflexión de Bragg es un fenómeno característico de la propagación de una onda en un cristal. Asimismo, la reflexión de Bragg es la causante de la aparición de las llamadas bandas prohibidas de energía.

La primera banda de energía se asocia con la primera reflexión de Bragg, que corresponde a la siguiente expresión:

$$n\lambda = 2d \sin\theta \quad [1]$$

Que para incidencia normal y n=1, queda:

$$\lambda = 2d = \frac{c}{f} \rightarrow f = \frac{c}{2d} = \frac{c}{2a} \rightarrow k = \frac{\pi}{a} \quad [2]$$

Que corresponde con el límite de la zona de Brillouin. Esta expresión es equivalente a la de condición de difracción:

$$(k')^2 = (k+G)^2 \quad [3]$$

Que podemos simplificar a la siguiente expresión:

$$k = \pm \frac{1}{2}G \quad [4]$$

Por lo tanto:

$$k = \pm \frac{1}{2}G = \pm n\frac{\pi}{a} \quad [5] \text{ (Meseguer et al., 1999)}$$

Así, la primera reflexión de Bragg ocurre para $k = \pm \frac{\pi}{a}$, es decir, para aquel vector que coincide con el vector que queda dentro del límite de la primera zona de Brillouin y, por tanto, cumple las condiciones de reflexión.

La reflexión aparece para ese valor debido a que la onda reflejada por un átomo en la red lineal interfiere constructivamente con la onda reflejada por el átomo vecino más próximo. La diferencia de fase entre las dos ondas reflejadas es $\pm 2\pi$.

Se puede demostrar que para $k = \pm \frac{\pi}{a}$ las funciones de onda no son las ondas de propagación $e^{\frac{i\pi x}{a}}$ y $e^{\frac{-i\pi x}{a}}$ del modelo de electrones libres, sino que las soluciones para estos valores particulares de k están formados igualmente por ondas que viajan hacia la derecha y hacia la izquierda: las soluciones son ondas estacionarias.

Se puede afirmar entonces que cuando se satisface la condición de Bragg, una onda no se puede propagar en la red, estableciéndose una onda estacionaria por reflexiones sucesivas, por lo tanto ha aparecido una banda prohibida de propagación.

Existe una banda prohibida en $k = \pm\frac{\pi}{a}$, que corresponde con la primera reflexión de Bragg y con los límites de la primera zona de Brillouin. Las siguientes zonas prohibidas se encuentran en $\pm n\frac{\pi}{a}$.

## 3. METODOLOGÍA

El proceso de análisis se puede esquematizar en los siguientes pasos:

a) En una primera fase, se procede a un estudio acústico de la piscina en la cual son sumergidas posteriormente las estructuras, se trata principalmente de establecer la pareja de transductores más apta para las medidas a realizar, así como un estudio de la ubicación idónea de los mismos.

b) Se elige la señal más apta para ser usada y se emite en un rango de frecuencias ultrasónicas con la piscina sin estructuras, a fin de obtener unos resultados que nos sirvan a posteriori como referencia.

c) Se eligen las estructuras periódicas a estudio y se calculan teóricamente aquellas frecuencias que serán susceptibles de presentar atenuación.

d) Se sumergen las estructuras periódicas y se comparan los resultados obtenidos con los de referencia, de esta manera calculamos la variación sufrida por la señal debida únicamente a la disposición de la estructura y no a otras pérdidas.

A continuación se detallan los procesos seguidos en la instalación y configuración de la estructura, así como de la obtención de la señal de referencia y la disposición de los materiales bajo estudio.

### 3.1. Montaje y equipo de medida

Dado que el objetivo es estudiar estructuras periódicas que favorezcan la aparición del fenómeno de bandas prohibidas y permitidas en la zona ultrasónica, la principal preocupación es, por una parte, la correcta construcción y ubicación de las estructuras en el seno del elemento usado como medio de propagación (material huésped) y, por otra, la elección del método de medida más adecuado para las mediciones.

### 3.1.1. Montaje de las estructuras

Las estructuras se sumergen en una piscina de dimensiones 87.5 x 113 x 56.5 cm., teniendo en cuenta 1.5 cm. de grosor en los límites laterales y en el fondo de la misma.

Para la sujeción de los cilindros se usa una plataforma circular de madera de 54 cm. de diámetro y 1.6 cm. de grosor; en ella se practican agujeros del diámetro correspondiente (una plataforma por cada diámetro) de forma que se puedan montar todas las redes a tratar.

A cada uno de los cilindros se les suelda un tope en uno de sus extremos, de manera que al introducir los cilindros a través de los agujeros correspondientes de la madera, éstos queden suspendidos sobre ésta.

La plataforma de madera se suelda a una estructura que le permite rotar sobre su propio eje de manera que la estructura cristalina montada pueda girar y estudiar de ese modo el efecto en función de la dirección de incidencia, con lo que además los transductores pueden mantenerse fijos en sus posiciones originales durante todas las mediciones.

La piscina se llena de agua hasta alcanzar una altura de 40 cm. (lo que equivale aproximadamente a un volumen de 370 litros). De esta manera, como los cilindros tienen una longitud de 45 cm., éstos sobresalen por encima de la superficie del agua logrando que, cuando la onda se propague, ésta sufra dispersión a lo largo de todo el eje OZ.

La piscina se deposita sobre una gran estructura de metal y, para evitar el contacto directo de ésta con el cristal, se coloca una capa de poliéster, la cual además sirve para amortiguar la fijación de la misma.

Como medida adicional de seguridad se fijan unas trinchas alrededor de la pecera (para proteger las paredes laterales de la presión ejercida por el volumen de agua sobre ellas) a una altura de 1/3 de la superficie del agua, zona donde la presión del agua es mayor.

Igualmente se colocan barras transversales, las cuales sirven para suspender tanto los transductores como la plataforma de madera sobre la superficie del agua. La sujeción a los bordes de la pecera se realiza mediante velcro, el cual permite una fijación que aísle las barras (de metal) del borde de la pecera sin dañar a ésta, y una sujeción mínima que evite movimientos laterales de las mismas.

El montaje final es mostrado parcialmente en las figuras 2 y 3:

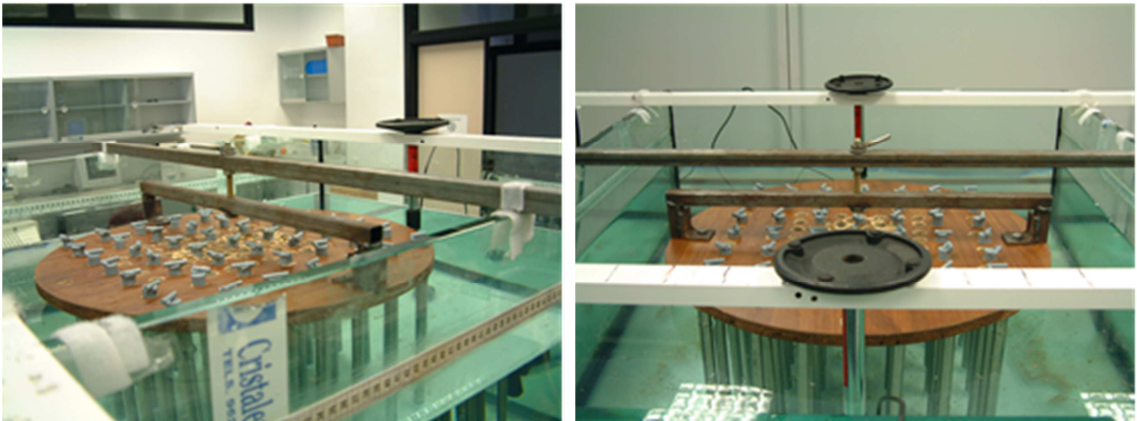

**Figuras 2 (izquierda) y 3 (derecha). Montaje de las estructuras en la piscina**

### 3.1.2. Método de medida

Con la intención de automatizar el proceso de medida, se intenta aplicar en un primer momento software *ad hoc* (Zurita, 2003), mediante el cual es posible generar señales continuas (*bursts*, MLS) y capturarlas. La intención es medir en primer lugar generando trenes de pulsos para posteriormente intentar medir mediante el uso de señales MLS.

Sin embargo, este procedimiento genera problemas en la captura de la señal; el programa es capaz de capturar el valor de pico a pico de la señal obtenida pero, en muchas ocasiones, este valor no se corresponde con la respuesta a la onda directa.

Este fenómeno es debido a que las reflexiones producidas por la superficie y el fondo de la piscina llegan a la vez al receptor, puesto que la distancia recorrida por cada onda es la misma. Al suceder esto, en ocasiones, las dos señales llegan en fase aumentando el nivel recibido, que a veces sobrepasa el valor pico a pico de la respuesta a la señal directa por lo que se obtienen valores equivocados.

Por ello, finalmente se opta por realizar las mediciones manualmente, con el fin de determinar para cada medida cuál es el valor pico a pico válido.

A partir de los trabajos realizados con anterioridad por Solano (2000), se elige como señal emisora un *burst* senoidal de 2 ciclos de 20 Vpp, pero con una frecuencia de repetición de *burst* bastante más baja (25 Hz), lo que permite una representación en la pantalla del osciloscopio más estable que para frecuencias de repetición mayores.

En primer lugar se emite esta señal con la piscina sin estructuras. Se realiza un barrido y se obtiene la respuesta en frecuencia de la piscina. Posteriormente se sumergen las estructuras y se vuelven a emitir las mismas señales, por lo que no queda más que comparar la señal recibida con y sin estructura y evaluar la pérdida o ganancia de nivel mediante la siguiente expresión (Martinez-Sala et al., 1995):

$$n = 20 log \frac{V_{pp}(inicial)}{V_{pp}(final)} \ (dB) [6]$$

En la figura 4 se muestra el esquema básico inicial del conexionado del equipo:

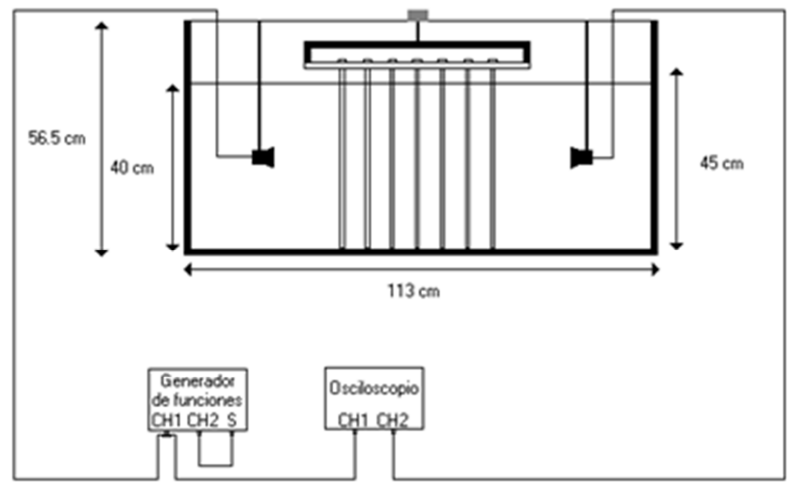

**Figura 4. Esquema del equipo de medida**

### 3.1.3. Caracterización acústica de la piscina

Seguidamente se procede a realizar un estudio acústico de la piscina con el fin de conocer su comportamiento frente a la propagación de las señales emitidas. Se debe elegir la pareja de transductores más adecuada y ubicarlos en el mejor emplazamiento posible para evitar cualquier contaminación de las medidas.

Una vez elegidos los transductores a usar y su disposición, se debe obtener la señal de referencia con la que comparar los resultados que se obtengan con las estructuras sumergidas.

*Profundidad de trabajo*

Un receptor colocado en las proximidades de un emisor va a recibir señal directa y reflexiones, tanto de las paredes laterales como de la superficie y el fondo, que no llegarán a la vez debido a la diferencia de longitud de los caminos recorridos.

La reflexión de las ondas puede llegar a combinarse con la onda directa y, en función de sus fases, puede producir una amplitud mayor o menor en la presión original de la onda directa. Debido a esto, es importante colocar los transductores a una profundidad tal que las reflexiones tanto en la superficie como en el fondo estén lo más atenuadas posibles.

### a) superficie

Las pérdidas en la superficie dependen de:
- La frecuencia de radiación.
- El ángulo de incidencia.
- La rugosidad en la superficie:

    - Si la superficie del agua está completamente lisa, la reflexión será casi perfecta y el índice de reflexión será prácticamente "1", produciéndose una onda reflejada de gran intensidad y con un cambio de fase de 180º (Solano, 2000).

    - Por el contrario, si la superficie es rugosa debido a la acción del viento y el oleaje en el caso del mar o gracias a un motor en el caso de una pecera, la reflexión se dispersa y por tanto se atenúa.

### b) suelo

Las reflexiones en el fondo de la piscina suelen ser difíciles de evaluar, pero son de una energía menor que las debidas a las reflexiones en la superficie.

La ubicación de los transductores debería ser a 2/3 de la profundidad total pues las reflexiones debidas a la superficie son más fuertes (Kutruff, 1991), pero en el caso de tanques o peceras de reducidas dimensiones, deben ser colocarlos justo a la mitad. En la figura 5 se puede observar el gráfico de la disposición de la plataforma sumergida a una profundidad de 20 cm.

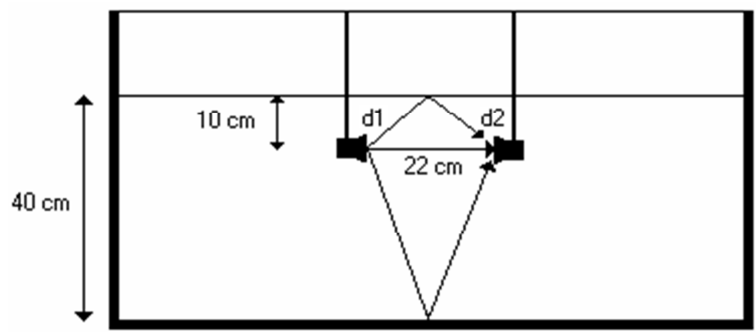

**Figura 5. Plataforma y esquema de reflexiones a 20 cm**

Por lo general, la velocidad del sonido en un líquido disminuye al aumentar la temperatura, pero en el caso concreto del agua dulce se ha podido comprobar experimentalmente que la velocidad aumenta con la temperatura hasta un valor límite (aproximadamente 73ºC), por encima de este punto hay una disminución de la velocidad, como se puede observar en la figura 6:

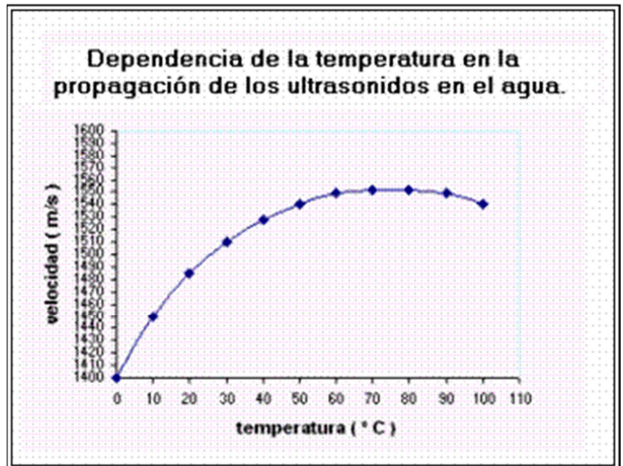

**Figura 6. Dependencia de la temperatura en la propagación de los ultrasonidos en agua**

Tras medir la temperatura del agua (20° C), se toma el valor de c= 1485 m/s

Conociendo la velocidad de propagación y las distancias relativas, el cálculo es trivial:

Teniendo en cuenta que ahora el retardo es igual tanto para la superficie como para el fondo, pues el recorrido es el mismo, obtenemos los siguientes valores:

$$t = \frac{e}{c} = \frac{d1+d2}{1485} = 3 \cdot 10^{-4} s \ [7]$$

$$d1 = d2 = \sqrt{(0.2^2) + (0.11^2)} \cong 0.228 m \ [8]$$

La señal recibida se puede observar en la figura 7:

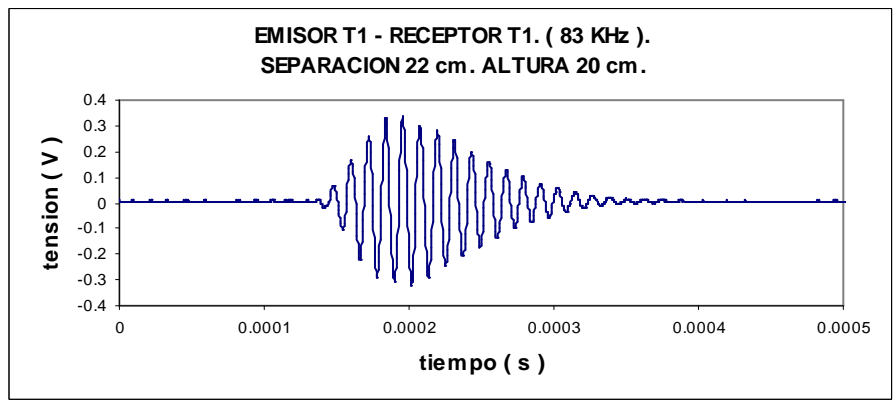

**Figura 7. Señal recibida (83KHz), a una separación de 22cm y profundidad de 20cm**

Tanto la reflexión debida a la superficie como al fondo llegan a la vez y además cuando la onda directa casi ha desaparecido, por lo que no van a afectar en principio al valor de pico máximo, que es el que interesa para tomar referencias.

Se debe tener presente que la distancia de separación no es la definitiva, es una distancia útil para determinar cómo afectan las reflexiones del suelo y la superficie y saber a qué profundidad su efecto es menor. De hecho, la distancia de separación definitiva es de hecho el factor que falta por determinar para conocer la ubicación final de los transductores.

*Separación entre los transductores*
Se debe elegir aquella distancia para la cual la respuesta de la piscina sea más lineal y además el nivel de respuesta sea lo suficientemente alto. Se debe tener igualmente presente que, pese a que las siguientes configuraciones se estudian hasta cerca de los 450 kHz, la frecuencia de trabajo será hasta los 100 kHz y, por tanto, a partir de esa frecuencia no importa cómo se comporte la piscina.

En este caso, la distancia de separación se calculó para tres valores diferentes (57, 67 y 77 cm.). Las respuestas fueron muy parecidas, no obstante se descartó la de 57 cm. debido a que, cuando las estructuras quedaran montadas, éstas quedarían a tan sólo un par de centímetros de los transductores, lo cual podría ocasionar problemas. Respecto a la distancia de 77 cm; el problema consiste en que los transductores quedan muy cerca físicamente de las paredes delanteras y traseras de la piscina, y los rebotes en dichas zonas podrían afectar. Por ello, la distancia elegida finalmente es la de 67 cm, (figura 8).

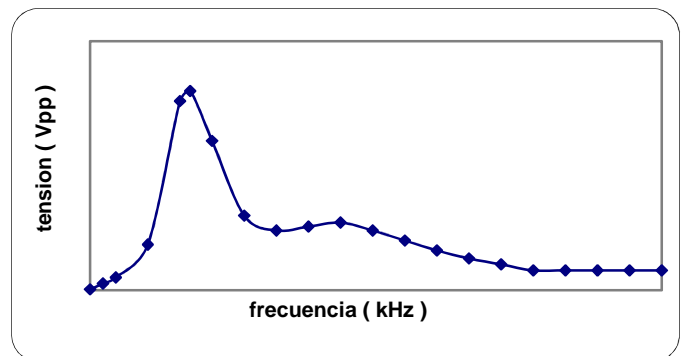

**Figura 8. Barrido de señal para una separación entre transductores de 67 cm**

*Paredes laterales*
Al igual que ocurre con la superficie, cuando las ondas acústicas inciden sobre las paredes laterales, se producen ondas reflejadas que quedarán registradas por el transductor y podrán alterar las medidas. El efecto de las reflexiones en las paredes laterales se puede observar en la señal mostrada en la figura 9.

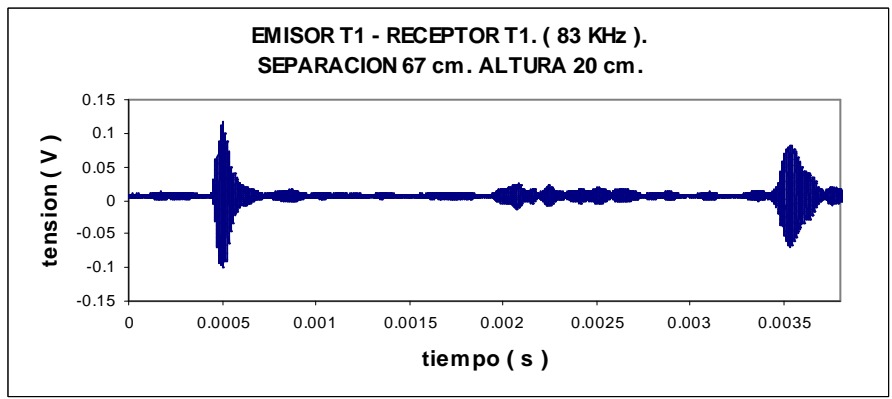

**Figura 9. Efecto de reflejos producidos en paredes laterales (separación 67 cm.)**

Con el propósito de catalogar cada reflexión, se realiza el siguiente análisis de rayos:

**Rayo directo:**

$t = \frac{e}{c} = \frac{67 \cdot 10^{-2}}{1485} = 4.51 \cdot 10^{-4} s$ [9]

**Reflexión superficie y suelo:**

$$t = \frac{e}{c} = \frac{d1+d2}{1485} = \frac{0.78}{1485} = 5.25 \cdot 10^{-4} s \quad [10]$$

$$d1 = d2 = \sqrt{(0.20^2) + (0.335^2)} \cong 0.39 m \quad [11]$$

**Reflexión en la pared lateral:**

$$t = \frac{e}{c} = \frac{d1+d2}{1485} = \frac{1.0783}{1485} = 7.26 \cdot 10^{-4} s \quad [12]$$

$$d1 = d2 = \sqrt{(0.4225^2) + (0.335^2)} \cong 0.539 m \quad [13]$$

**Reflexión en pared trasera + delantera:**

$$t = \frac{e}{c} = \frac{2.87}{1485} = 1.93 \cdot 10^{-3} s \quad [14]$$

$$e = 0.67 + 0.215 + 1.1 + 0.15 + 0.67 = 2.87 m \quad [15]$$

En principio se debe eliminar toda señal que no sea la directa, por ello se intenta atenuar todas las reflexiones mediante la colocación de diverso material absorbente en las paredes, pero no se consiguen resultados favorables. En este tipo de mediciones suele ser normal el uso de INSULKRETE (Darner, 1954), material formado mediante la mezcla de cemento y serrín, pero dado que no se dispone de dicho material se decide no colocar absorbente y controlar la distancia.

    Sin embargo, las reflexiones en el fondo y en la superficie sí van a ser conflictivas pues llegan bastante tempranas. El efecto se podría minimizar si las dimensiones de la piscina fuesen mayores y se dispusiera de más centímetros de profundidad, pero como no es así se debe tener precaución a la hora de tomar el valor pico a pico de respuesta directa, y no confundirlo con una reflexión.

*La señal de referencia*
Una vez elegidos los transductores y establecida tanto la profundidad de los mismos como su separación, el siguiente paso es obtener la señal de referencia. Para ello se realiza un barrido en frecuencia más profundo con la configuración elegida en el apartado anterior, cuyos resultados se pueden observar en la figura 10.

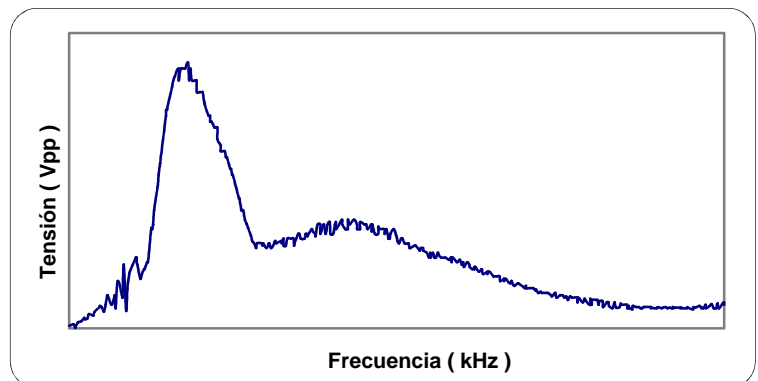

**Figura 10. Medida de referencia (piscina vacía, sin amplificación)**

El principal problema de esta señal es que es muy irregular desde los 30 kHz a los 50 kHz, y además los niveles son relativamente bajos. Se debe tener presente que se están emitiendo 20 Vpp y, en tan sólo 67 cm., llegan apenas 0.22 Vpp. Además, cuando las estructuras sean sumergidas, los niveles se atenuarán todavía más y será muy difícil obtener niveles adecuados de respuesta.

Debido a esto, se decide usar un amplificador de tensión a la salida del sistema. La ganancia de salida se coloca al máximo posible (32 dB), 10 KΩ de resistencia de entrada y una capacidad de entrada de 3.3 nF. Ahora, el equipo de medida queda de la siguiente forma (figura 11):

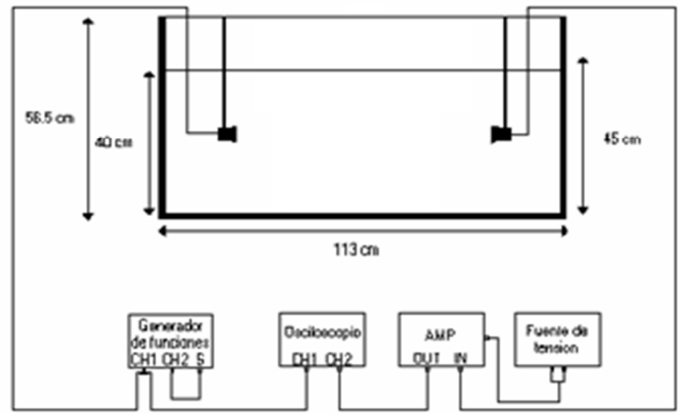

**Figura 11. Esquema final de medida (con amplificador)**

Tras realizar un nuevo barrido en frecuencia con el amplificador, se obtiene la siguiente respuesta (figura 12):

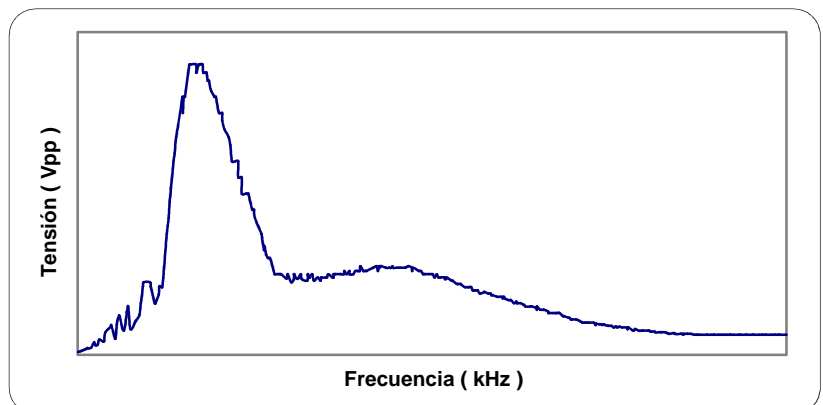

**Figura 12. Medida de referencia (piscina vacía, con amplificación)**

Variando los parámetros de resistencia de entrada y de capacidad, se pueden conseguir niveles de ganancia aún mayores, pero la distorsión de la señal también es mayor por lo que se opta por los valores elegidos.

Los huecos en frecuencia que se pretenden observar poseen anchos de banda muy pequeños; en medio aéreo, estos anchos no sobrepasan los 500 Hz. Debido a esto es necesario realizar un barrido muy profundo para no pasar por alto algún hueco en frecuencia.

Por otra parte los huecos se van a repetir tantas veces como repeticiones tenga la estructura, es decir, una red cuadrada de 7x7 deberá poseer teóricamente 7 huecos en frecuencia, por lo que el estudio hasta los 450 kHz como hasta ahora hemos hecho resulta inviable e innecesario. Debido a esto la frecuencia máxima de trabajo se fija en 100 kHz. Por tanto, la frecuencia de referencia final se muestra en la figura 13.

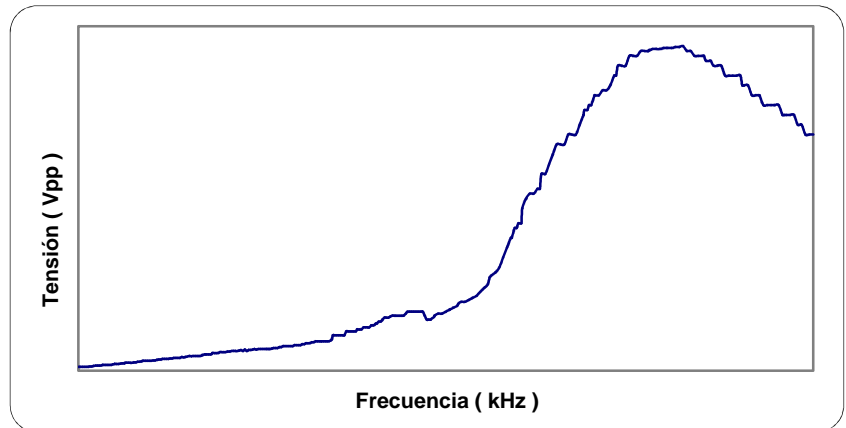

**Figura 13. Medida de referencia final**

### 3.2. Material huésped y dispersor

El elemento dispersor elegido ha sido acero inoxidable y el medio huésped agua dulce, de manera que intentaremos formar una serie de estructuras en 2 dimensiones que simulen a las de un cristal y las sumergiremos en agua.

El material elegido para los cilindros ha sido acero inoxidable por varias razones: por una parte dada la necesidad de estar sumergidos era conveniente que el material no se oxidara y por otra el proceso de dispersión se verá favorecido dada la diferencia de impedancia entre el agua y el acero.

El cristal de sonido en 2 dimensiones va a presentar una periodicidad de densidades a lo largo de los ejes OX y OY mientras que va a ser homogéneo a lo largo del eje OZ. Los materiales dispersores serán cilindros de manera que la estructura cristalina estará formada por estos cilindros de sección circular dispuestos en diferentes mallas.

Debido a que las muestras realizadas van a ser en 2-D es necesario que éstas sean de una longitud lo más larga posible. En nuestro caso este factor viene limitado por la altura de la piscina, pero debido a que es necesario montar una estructura para suspender los cilindros y permitir su giro, con el orden de estudiar la atenuación en diferentes direcciones de simetría, los cilindros serán piezas de 45 cm de longitud de acero inoxidable, de densidad de 7900 kg/m$^3$.

En las figuras 14 y 15 podemos observar la disposición periódica de los cilindros para el caso de una red cuadrada:

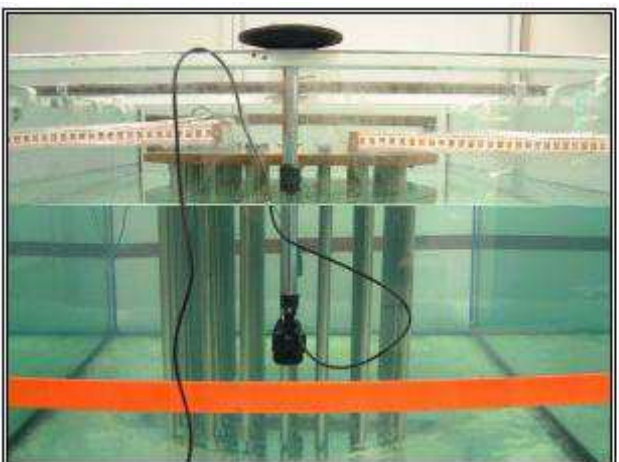
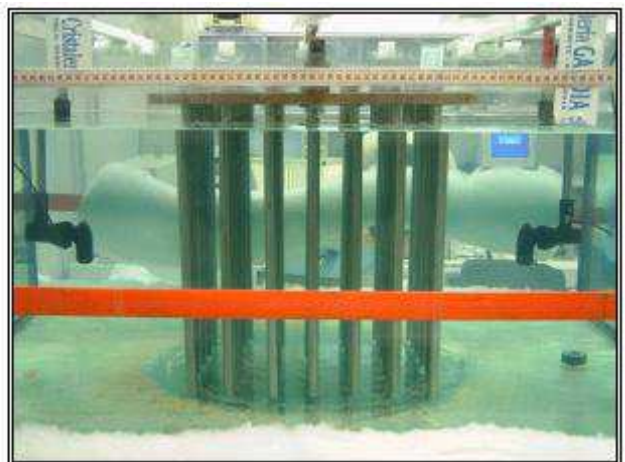

**Figuras 14 y 15. Diferentes vistas de la piscina**

Uno de los factores que intervienen en la aparición de las bandas prohibidas de propagación es el factor de llenado, y éste depende de la periodicidad de la red, del tipo de red y del diámetro de los centros dispersores; es decir, del diámetro de los cilindros.

Por ello, y con el orden de estudiar el comportamiento para distintos factores de llenado, se realizan medidas para dos tamaños distintos de cilindros (8 y 16 mm de diámetro).

Las medidas se realizan tanto para cilindros macizos como huecos (de 1 mm de grosor) con el fin de estudiar las posibles diferencias que puedan surgir al variar estos parámetros, y en distintas en redes: cuadrada, triangular y hexagonal. Las medidas completas realizadas en esta investigación se observan en la tabla 1.

| TIPO DE RED | PERIODICIDAD | DIÁMETRO | ÁNGULO DE INCIDENCIA |
|---|---|---|---|
| **CUADRADA** | 30 mm, 60 mm | 8 mm, 16 mm | 0°, 15°, 30°, 45°, 60°, 75° |
| **CUADRADA REDUCIDA** | 60 mm | 16 mm | 0°, 15°, 30°, 45°, 60°, 75° |
| **CUADRADA (EFECTO BORDE )** | 60 mm | 16 mm | 0°, 15°, 30°, 45°, 60°, 75° |
| **TRIANGULAR** | 67.08 mm | 8 mm, 16 mm | 0°, 15°, 30°, 45° |
| **HEXAGONAL** | 120 mm | 8 mm, 16 mm | 0°, 15°, 30°, 45° |
| **HEXAGONAL REDUCIDA** | 120 mm | 16 mm | 0°, 15°, 30°, 45° |

**Tabla 1. Esquema general de mediciones: red, periodicidad, diámetro y ángulo de incidencia**

En este trabajo se muestran únicamente los resultados obtenidos para las redes triangulares (en grados de incidencia 0º y 30º), mientras que los datos para el resto de redes serán tratados en futuros estudios.

## 4. RESULTADOS

### 4.1. Estructura con base triangular

En la figura 16 se representan gráficamente las estructuras creadas con los cilindros para simular la base triangular, mostrando igualmente las principales zonas de simetría.

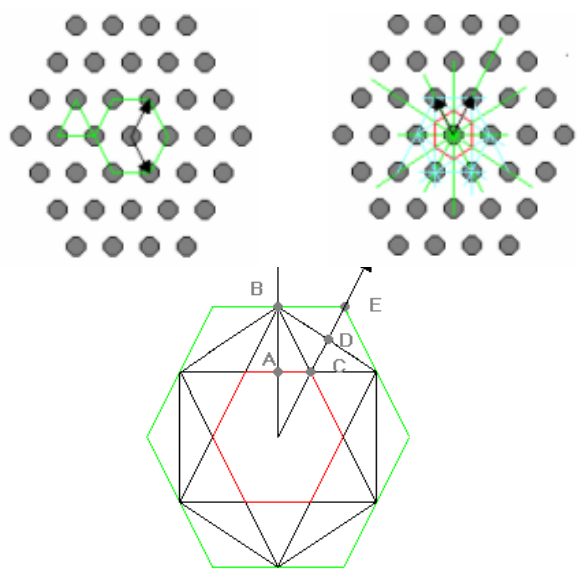

**Figura 16. Red directa (superior izquierda) y recíproca (superior derecha), y primeras zonas de Brillouin (inferior). Principales zonas de simetría (0º y 30º)**

### 4.1.1. Predicción de teórica

En la tabla 2 se muestran las predicciones teóricas de supuesta aparición de zonas de atenuación:

| Periodicidad (mm) | Vectores Directos (mm) | Vectores recíprocos (mm$^{-1}$) | K (mm$^{-1}$) | | λ (mm$^{-1}$) | Frecuencia atenuación ( Hz ) |
|---|---|---|---|---|---|---|
| 67,08 | $\vec{a} = 67.08\vec{\imath}$ $\vec{b} = 33.54\vec{\imath} + 60\vec{\jmath}$ $\vec{c} = \vec{k}$ | $\vec{a} = 0.09\vec{\imath} - 0.05\vec{\jmath}$ $\vec{b} = \frac{2\pi}{60\vec{\jmath}}$ $\vec{c} = 2\pi\vec{k}$ | OA | 0.052 | 120.000 | 12375.0 |
| | | | OB | 0.105 | 60.000 | 24750.0 |
| | | | OC | 0.060 | 103.923 | 14289.4 |
| | | | OD | 0.091 | 69.282 | 21434.1 |
| | | | OE | 0.121 | 51.962 | 28578.8 |

**Tabla 2. Predicción teórica de zonas de atenuación**

Para los cálculos usados, se han tenido en cuenta las siguientes relaciones de geometría:

$OC = OA \frac{2}{\sqrt{3}}$ [16]

$OD = OE \frac{3}{4}$ [17]

En el caso particular de una red triangular, las direcciones principales de simetría son las correspondientes a 0° y 30°. Pese a que en toda la bibliografía existente acerca del tema son usados los índices de Miller para definir direcciones de incidencia, a partir de ahora la dirección quedará determinada únicamente en grados para facilitar la comprensión (tabla 3).

| (dirección 0°) | (dirección 30°) |
|---|---|
| 12375 Hz | 14289.4 Hz |
| 24750 Hz | 28578.8 Hz |
| 37125 Hz | 42868.2 Hz |
| 49500 Hz | 57157.6 Hz |
| 61875 Hz | 71447 Hz |
| 74250 Hz | 85736.4 Hz |
| 86625 Hz | |

**Tabla 3. Frecuencias de atenuación por ángulo de incidencia (0º y 30º)**

### 4.1.2. Factor de llenado

En este caso, el área total de la composición es el área del triángulo, mientras que el área ocupada por los centros dispersores es $\frac{\pi r^2}{2}$ (figura 17).

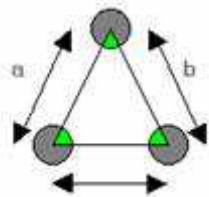

**Figura 17. Factor de llenado (estructura triangular)**

El factor de relleno resultante es ***f.f.*:** $\frac{2\pi r^2}{\sqrt{3}a^2}$ (Kushwaha, 1997), los valores correspondientes se muestran en la tabla 4:

| Cte. de red (mm) | Radio (mm) | Filling Fraction |
|---|---|---|
| 67.082 | 8 | 0.0515 |
| 67.082 | 16 | 0.2063 |

**Tabla 4. Factor de llenado**

### 4.1.3. Medidas para una red triangular (67.08 mm)

*Cilindros huecos (8 mm)*
Las figuras 18 y 19 muestran los resultados obtenidos en cilindros huecos de 8 mm, para una incidencia normal y a 30º respectivamente. Pese a presentar respuestas muy poco atenuadas, los huecos frecuenciales aparecen con bastante facilidad. En el caso de incidencia normal, se presenta la respuesta con menor nivel, llegando a marcar 9.11 dB en el hueco correspondiente a 47.5 kHz, valor más bajo obtenido. Respuesta muy oscilante y sin apenas atenuación. A los 36 kHz la respuesta presenta un pico de -0.5 dB de atenuación, y todos los huecos frecuenciales aparecen ligeramente desplazados respecto a los teóricos.

Por otro lado, se obtiene un hueco total (47.5 kHz) para todas las direcciones de incidencia, por lo que se podría considerar que es generado por el recinto de medida y no por la topología de la red.

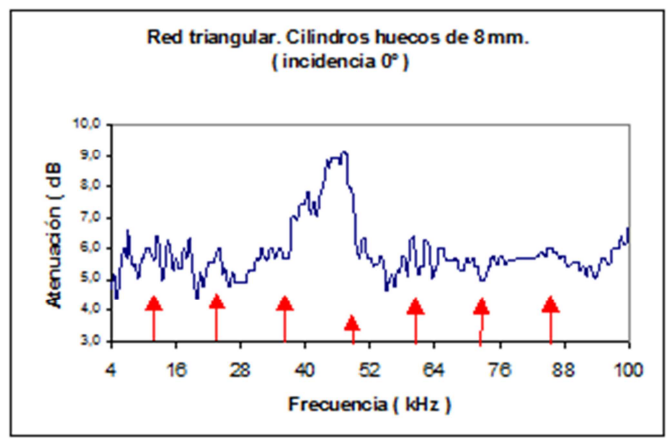

**Figura 18. Red triangular (cilindros huecos- 8mm; incidencia 0º**

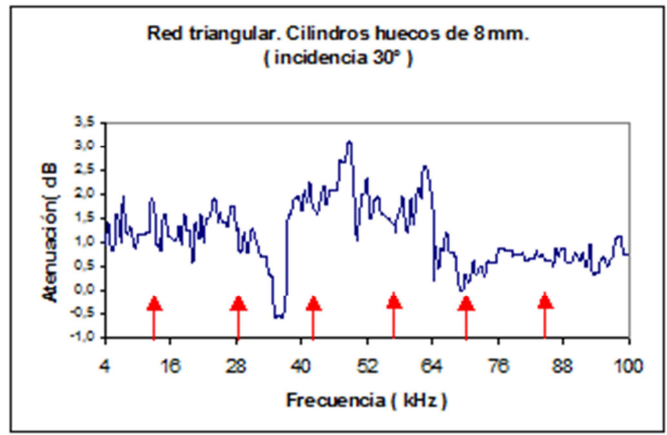

**Figura 19. Red triangular (cilindros huecos; 8 mm; incidencia 30º)**

*Cilindros huecos (16 mm)*

Las figuras 20 y 21 muestran por su parte los valores obtenidos con el conjunto de cilindros de mayor diámetro (16 mm), tanto en incidencia normal como a 30º.

Respecto a la incidencia normal, excepto el hueco predicho para los 86625 Hz, los demás coinciden plenamente. La respuesta es muy débil llegando a los 29.3 dB de atenuación para el primer hueco frecuencial a los 12375 Hz,y a partir de los 78 kHz la respuesta comienza a crecer poco a poco hasta los 100 kHz.

Respecto a la incidencia a los 30º, Respuesta muy aleatoria, todos los huecos frecuenciales teóricos se obtienen. La respuesta obtenida no está tan fuertemente atenuada como ocurría para las anteriores medidas, y aparece un hueco a los 23 kHz bastante importante el cual no había sido predicho.

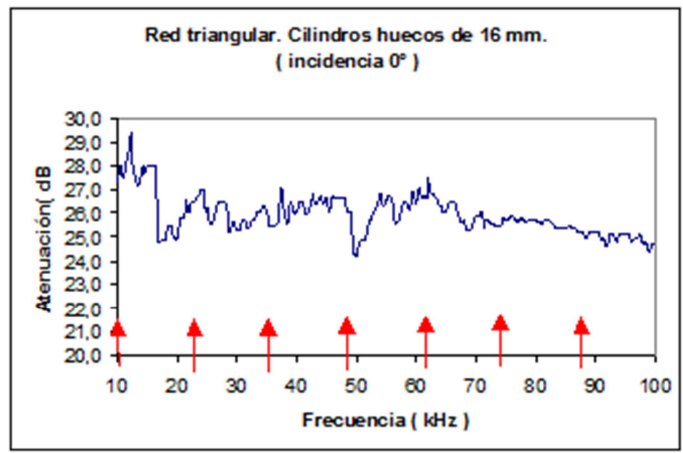

**Figura 20. Red triangular (cilindros huecos; 16 mm; incidencia 0º)**

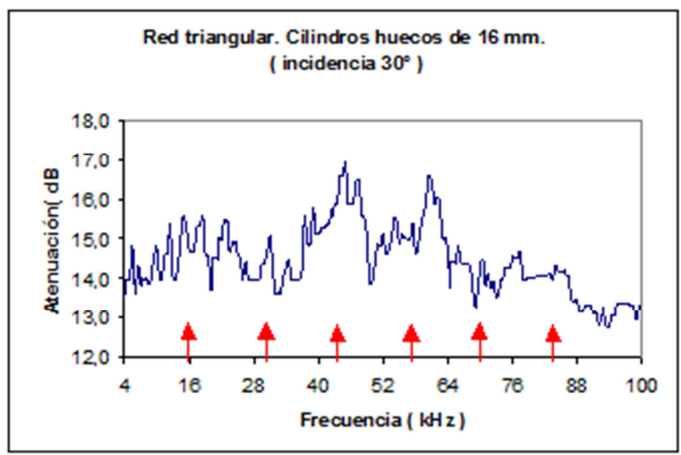

**Figura 21. Red triangular (cilindros huecos; 16 mm; incidencia 30º)**

*Cilindros macizos (8 mm)*

Las respuestas correspondientes a los cilindros macizos presentan unos niveles muy parecidos, excepto para el caso de la incidencia a 0° (figura 22), donde se presentan atenuaciones mayores que a 30º de incidencia (figura 23). En estas dos direcciones principales de simetría se consiguen resultados de gran similitud con las predicciones.

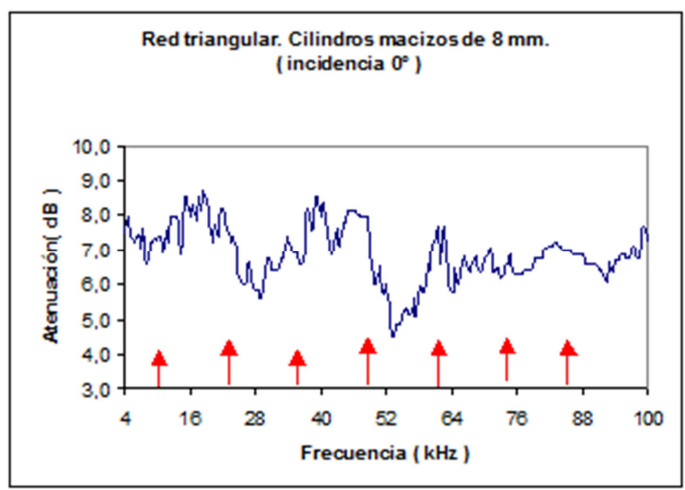

**Figura 22. Red triangular (cilindros huecos; 16 mm; incidencia 30º)**

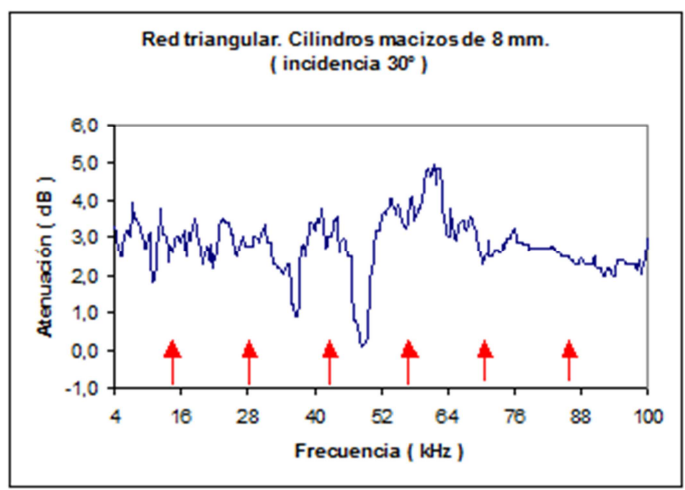

**Figura 23. Red triangular (cilindros huecos; 16 mm; incidencia 30º)**

*Cilindros macizos (16 mm)*
Finalmente, las figuras 24 y 25 muestran los resultados obtenidos con la muestra de cilindros macizos de diámetro mayor (16mm).

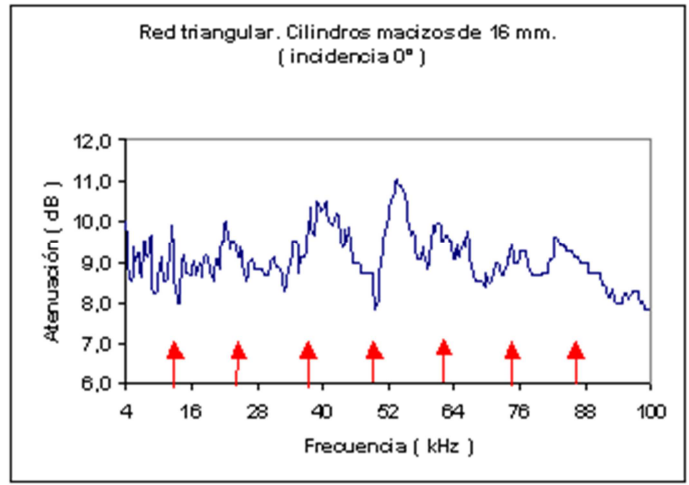

**Figura 24. Red triangular (cilindros huecos; 16 mm; incidencia 30º)**

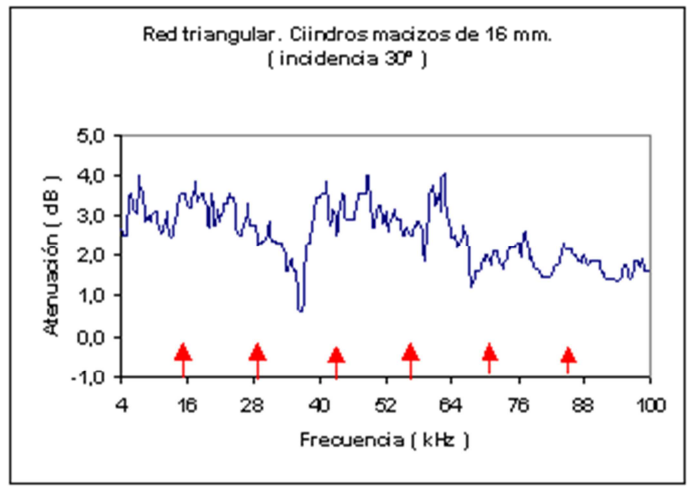

**Figura 25. Red triangular (cilindros macizos; 16 mm; incidencia 30º)**

### 4.1.4. Comparativa entre las distintas configuraciones.

Finalmente, en la figura 26 se muestra una comparativa general de los resultados en función del diámetro de los cilindros, del grado de incidencia (0º y 30º) y de la composición (huecos o macizos).

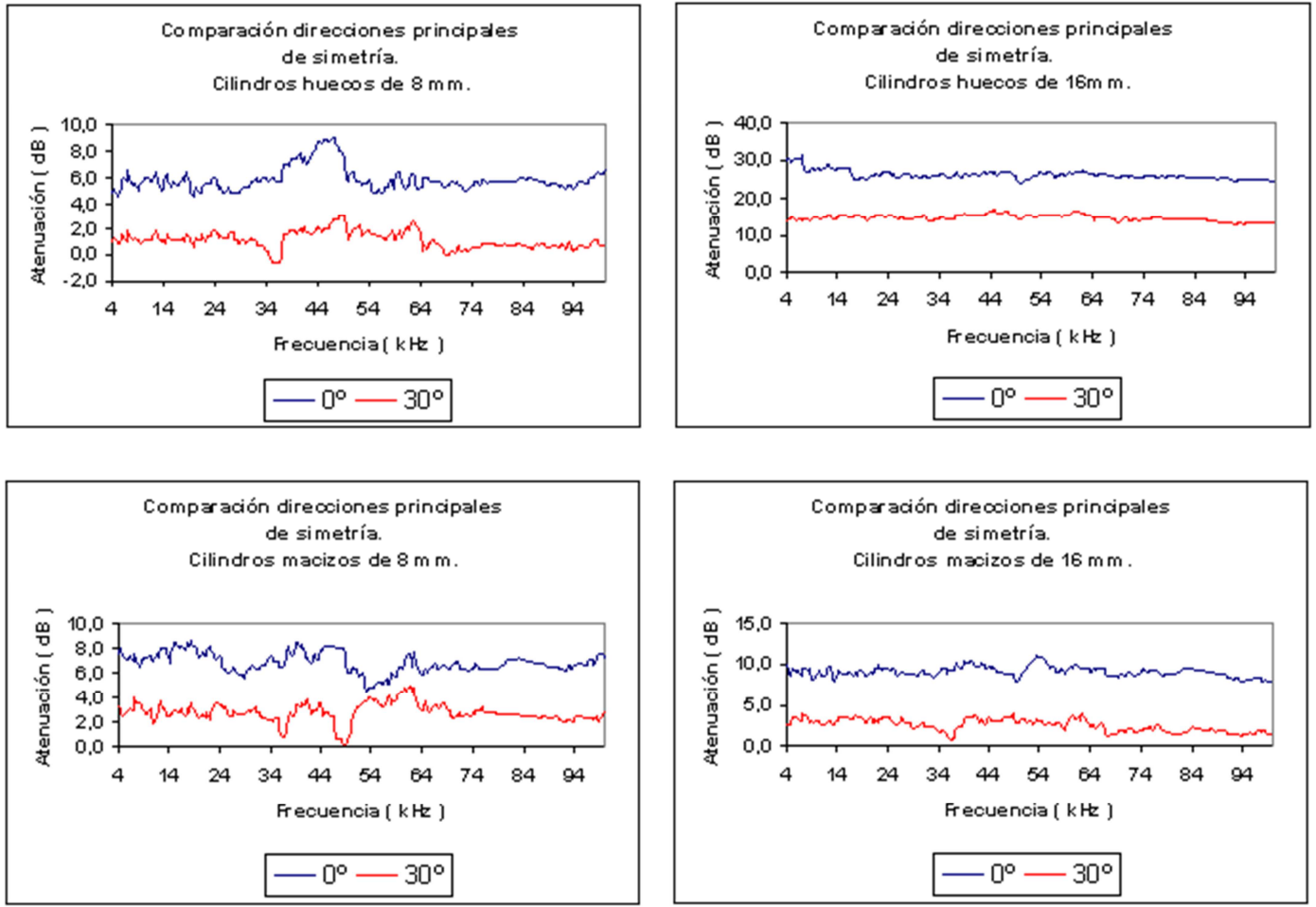

**Figura 26. Comparativa en función del diámetro, composición y grado de incidencia**

## 5. DISCUSIÓN Y CONCLUSIONES

A lo largo de este trabajo se han analizado los distintos patrones de dispersión generados por estructuras periódicas sumergidas en medio líquido con el fin de determinar sus propiedades frente a la transmisión de haces ultrasónicos a su través.

Se ha conseguido probar en modelos a escala la atenuación en ciertas bandas de frecuencia en función de las características geométricas de la estructura. Esto implica que se puede inhibir la propagación del sonido a ciertas frecuencias, funcionando las estructuras como filtros selectivos en frecuencia. Estos resultados concuerdan parcialmente con los obtenidos por Ye y Hoskinson (2000).

Las principales conclusiones obtenidas son las siguientes:

- La configuración triangular presenta zonas de atenuación selectivas (huecos frecuenciales). Además se obtienen unos resultados parecidos a los predichos teóricamente para las dos direcciones principales de simetría.
- Empleando cilindros macizos, apenas hay diferencias entre 8 y 16 mm de diámetro, pero usando cilindros huecos (1mm de grosor), las diferencias sí son grandes, recibiendo en el caso de cilindros de 8 mm señales mucho más elevadas.
- Para cilindros de 8 mm de diámetro apenas se observan cambios importantes al variar de cilindros huecos a macizos, pero en cambio usando cilindros de 16 mm sí aparecen más diferencias.
- Centrado aproximadamente a los 44 kHz, aparece en casi la totalidad de las medidas un hueco frecuencial de dimensiones importantes, este hueco es independiente de la dirección de incidencia y de la red tratada. Las causas de su aparición pueden ser debidas a una posible contaminación de los niveles de señal directa por las reflexiones tanto en fondo como en la superficie, que a esa gama de frecuencias se solapen con la señal directa y disminuyan el valor de esta última.

Las propiedades de los cristales de sonido han comenzado a ser estudiadas relativamente hace poco tiempo, hay estudios importantes en medio gaseoso (transmisión en el aire) y en medio sólido (transmisión por la corteza terrestre), pero en medio líquido apenas hay información al respecto. Por lo que este trabajo supone una novedad.

Las aplicaciones de los cristales de sonido a la acústica subacuática pueden ser muy de interés. Se debe tener en cuenta que la estructura no va transmitir las frecuencias que se deseen atenuar mediante una construcción adecuada, por lo que su uso como pantallas acústicas es inevitable. Dado un recinto de medida subacuático y conociendo de antemano la frecuencia de ruido ambiente, se podría delimitar la zona de trabajo mediante paredes formadas por estructuras de este tipo eliminando de esta forma todo el ruido ambiente en las medidas.

De la misma forma que podemos aislar el interior de un recinto rodeado de estructuras periódicas del ruido exterior, también se podría aislar el exterior del posible ruido que se generase dentro de este recinto.

Además se podría disponer de varias estructuras en cascada, cada una de ellas diseñada para eliminar unas frecuencias de modo que el rango total inhibido fuese más amplio. De hecho, el estudio de la disposición de estructuras periódicas en cascada podría ser una línea de investigación interesante a seguir en un futuro.

Aparte de su uso como pantallas acústicas y todas las aplicaciones que eso lleva consigo, se ha demostrado que pueden ser elementos muy útiles para la localización e identificación de bancos de peces, entre otros elementos.

La disposición de los peces en un banco es idéntica a una estructura cristalina, es decir, forman estructuras periódicas. Es fácil llegar a la conclusión de que si supiésemos la respuesta de un sistema con unas características determinadas y la respuesta de un banco de peces se correspondiese con ésta, podríamos asemejar ambas formaciones e identificar el banco de peces con bastante fiabilidad.

El uso de cilindros huecos es muy importante pues al tener aire en su interior van a generar una analogía con el interior de los peces (ocupado parcialmente también de aire). De hecho, el estudio de cilindros huecos rellenos parcialmente de agua (y de cualquier otro elemento) queda pendiente para futuras investigaciones al respecto.

Como futuras líneas de investigación, aparte de las ya citadas, se podría añadir el estudio de otras estructuras periódicas en 2-D e incluso ampliar el trabajo a estructuras en 3-D, donde el número de redes posibles se ampliaría a 14, abriendo con ello un nuevo campo para la investigación.

En definitiva, la aplicabilidad de los cristales de sonido a la industria y a la investigación de fenómenos acústicos es muy elevada, los resultados prácticos se acercan bastante a los teóricos generando perspectivas muy prometedoras para el futuro.

## 6. REFERENCIAS BIBLIOGRÁFICAS